\documentclass[11pt]{article}
\usepackage{amssymb,cite}
\makeatletter
\@addtoreset{equation}{section}
\makeatother

\def\ben{\begin{equation}}
\def\een{\end{equation}}

  \let\n=\nu

\let\X=\Xi     
\let\C=\Chi

\def\nn{\nonumber} \def\bd{\begin{document}} \def\ed{\end{document}}
\def\ds{\documentstyle} \let\fr=\frac \let\bl=\bigl \let\br=\bigr
\let\Br=\Bigr \let\Bl=\Bigl
\let\bm=\bibitem
\let\na=\nabla
\let\pa=\partial \let\ov=\overline
\newcommand{\be}{\begin{equation}}
\newcommand{\ee}{\end{equation}}
\def\ba{\begin{array}}
\def\ea{\end{array}}
\def\ft#1#2{{\textstyle{{\scriptstyle #1}\over {\scriptstyle #2}}}}
\def\fft#1#2{{#1 \over #2}}
\def\del{\partial}
\def\vp{\varphi}
\def\sst#1{{\scriptscriptstyle #1}}
\def\oneone{\rlap 1\mkern4mu{\rm l}}
\def\td{\tilde}
\def\wtd{\widetilde}
\def\ie{\rm i.e.\ }
\def\dalemb#1#2{{\vbox{\hrule height .#2pt
        \hbox{\vrule width.#2pt height#1pt \kern#1pt
                \vrule width.#2pt}
        \hrule height.#2pt}}}
\def\square{\mathord{\dalemb{6.8}{7}\hbox{\hskip1pt}}}
\newcommand{\ho}[1]{$\, ^{#1}$}
\newcommand{\hoch}[1]{$\, ^{#1}$}
\newcommand{\bea}{\begin{eqnarray}}
\newcommand{\eea}{\end{eqnarray}}
\newcommand{\ra}{\rightarrow}
\newcommand{\lra}{\longrightarrow}
\newcommand{\Lra}{\Leftrightarrow}
\newcommand{\ap}{\alpha^\prime}
\newcommand{\bp}{\tilde \beta^\prime}
\newcommand{\tr}{{\rm tr} }
\newcommand{\Tr}{{\rm Tr} }
\def\0{{\sst{(0)}}}
\def\1{{\sst{(1)}}}
\def\2{{\sst{(2)}}}
\def\3{{\sst{(3)}}}
\def\4{{\sst{(4)}}}
\def\5{{\sst{(5)}}}
\def\6{{\sst{(6)}}}
\def\7{{\sst{(7)}}}
\def\8{{\sst{(8)}}}
\def\n{{\sst{(n)}}}
\def\cA{{{\cal A}}}
\def\cB{{{\cal B}}}
\def\cF{{{\cal F}}}
\def\cH{{{\cal H}}}
\def\tV{\widetilde V}
\def\tW{\widetilde W}
\def\tH{\widetilde H}
\def\tE{\widetilde E}
\def\tF{\widetilde F}
\def\tA{\widetilde A}
\def\im{{{\rm i}}}
\def\tY{{{\wtd Y}}}
\def\ep{{\epsilon}}
\def\vep{{\varepsilon}}
\def\R{\rlap{\rm I}\mkern3mu{\rm R}}
\def\bD{{{\bar D}}}
\def\R{\rlap{\rm I}\mkern3mu{\rm R}}
\def\bD{{{\bar D}}}
\def\R{{{\mathbb R}}}
\def\C{{{\mathbb C}}}
\def\H{{{\mathbb H}}}
\def\CP{{{\mathbb C}{\mathbb P}}}
\def\RP{{{\mathbb R}{\mathbb P}}}
\def\Z{{{\mathbb Z}}}
\def\bA{{{\mathbb A}}}
\def\bB{{{\mathbb B}}}
\def\bC{{{\mathbb C}}}
\def\bD{{{\mathbb D}}}
\def\bE{{{\mathbb E}}}
\def\bZ{{{\mathbb Z}}}
\def\Re{{{\mathfrak{Re}}}}
\def\Im{{{\mathfrak{Im}}}}
\def\cosec{{\,\hbox{cosec}\,}}
\def\Gm{{\Gamma_{\!\! -}}}
\def\Gp{{\Gamma_{\!\! +}}}
\def\stan{{standard }}
\def\nonstan{{supernumerary }}
\def\FF2{{ {}_{\sst 2}F_{\sst 1} }}
\def\FFF{{ {}_{\sst 3}F_{\sst 2} }}
\def\const{\rm constant}
\def\bog{{Bogomol'nyi\ }}

\newcommand{\tamphys}{\it Center for Theoretical Physics,
Texas A\&M University, College Station, TX 77843}

\newcommand{\upenn}{\it Department of Physics and Astronomy,\\ University
of Pennsylvania, Philadelphia, PA 19104}

\newcommand{\damtp}{\it DAMTP, Centre for Mathematical Sciences,
 Cambridge University,\\  Wilberforce Road, Cambridge CB3 OWA, UK}

\newcommand{\alberta}{\it Theoretical Physics Institute, 412 Physics 
Laboratory, University of Alberta, Edmonton, Canada}

\newcommand{\brussels}{\it Physique Th\'eorique et Math\'ematique,
Universit\'e Libre de Bruxelles,\\ Campus Plaine C.P. 231, B-1050
Bruxelles, Belgium}

\newcommand{\auth}{ M. Cveti\v c{\hoch *}, H. L\"u{\hoch {\ddagger}}, 
Don N. Page{\hoch {\dagger}} and C.N. Pope\hoch{\ddagger}}

\begin{document}
\begin{flushright}

UPR-1117-T\ \ \ MIFP-05-10\ \ \
Alberta Thy-07-05\\
{\bf hep-th/0505223}\\
May\  2005
\end{flushright}

\vspace{10pt}

\begin{center}

{\large {\bf New Einstein-Sasaki and Einstein Spaces from Kerr-de Sitter 
            }}

\vspace{20pt}
\auth

\vspace{20pt}

\hoch{*}\upenn

\vspace{10pt}

{\hoch{\ddagger}}{\it George P. \& Cynthia W. Mitchell
Institute for Fundamental Physics,\\ Texas A\& M University,
College Station, TX 77843-4242, USA}

\vspace{10pt}

\hoch{\dagger}\alberta

\vspace{20pt}

\underline{ABSTRACT}
\end{center}

   In this paper, which is an elaboration of our results in
hep-th/0504225, we construct new Einstein-Sasaki spaces
$L^{p,q,r_1,\cdots,r_{n-1}}$ in all odd dimensions $D=2n+1\ge 5$.
They arise by taking certain BPS limits of the Euclideanised Kerr-de
Sitter metrics.  This yields local Einstein-Sasaki metrics of
cohomogeneity $n$, with toric $U(1)^{n+1}$ principal orbits, and $n$
real non-trivial parameters.  By studying the structure of the
degenerate orbits we show that for appropriate choices of the
parameters, characterised by the $(n+1)$ coprime integers
$(p,q,r_1,\ldots,r_{n-1})$, the local metrics extend smoothly onto
complete and non-singular compact Einstein-Sasaki manifolds
$L^{p,q,r_1,\cdots,r_{n-1}}$.  We also construct new complete and
non-singular compact Einstein spaces $\Lambda^{p,q,r_1,\cdots,r_n}$ in
$D=2n+1$ that are not Sasakian, by choosing parameters appropriately
in the Euclideanised Kerr-de Sitter metrics when no BPS limit is
taken.





\pagebreak
\setcounter{page}{1}

\tableofcontents
\addtocontents{toc}{\protect\setcounter{tocdepth}{2}}
\newpage

\section{Introduction}

    Compact Einstein spaces have long been of interest in mathematics
and physics.  One of their principal applications in physics has been
in higher-dimensional supergravity, string theory and M-theory, where
they can provide backgrounds for reductions to lower-dimensional
spacetimes.  Of particular interest are Einstein spaces that admit
Killing spinors, since these can provide supersymmetric backgrounds.
In recent times, one of the most important applications of this type
has been to supersymmetric backgrounds AdS$_5\times K_5$ of type IIB
string theory, where $K_5$ is a compact Einstein space admitting
Killing spinors.  Such configurations provide examples for studying
the AdS/CFT Correspondence, which relates bulk properties of the
dimensionally-reduced AdS supergravity to a superconformal field
theory on the four-dimensional boundary of AdS$_5$
\cite{mald,guklpo,wit}.

    The most studied cases have been when $K_5$ is the standard round
5-sphere, which is associated with an ${\cal N}=4$ superconformal
boundary field theory.  Another case that has been extensively studied
is when $K_5$ is the space $T^{1,1}=(SU(2)\times SU(2))/ U(1)$.  Until
recently, these two homogeneous spaces were the only explicitly known
examples of five-dimensional Einstein spaces admitting Killing
spinors, although the general theory of Einstein-Sasaki spaces, which
are odd-dimensional Einstein spaces admitting a Killing spinor, was
well established, and some existence proofs were known (see, for
example, \cite{boygal1,boygal2}).

   In recent work by Gauntlett, Martelli, Sparks and Waldram, the
picture changed dramatically with the construction of infinitely many
explicit examples of Einstein-Sasaki spaces in five \cite{gamaspwa1}
and higher \cite{gamaspwa2} odd dimensions.  Their construction was
based on some earlier results in \cite{berber,pagpop}, in which local
Einstein-K\"ahler metrics of cohomogeneity 1 were obtained as line
bundles over Einstein-K\"ahler bases.  Using the well-known result
that an Einstein-Sasaki metric can be written as circle bundle over an
Einstein-K\"ahler metric, this yielded the new local metrics discussed
in \cite{gamaspwa1,gamaspwa2}.  In the case of five dimensions, the
resulting Einstein-Sasaki metrics are characterised by a non-trivial
real parameter.  They have cohomogeneity 1, with principal orbits
$SU(2)\times U(1)\times U(1)$.  In general these local metrics become
singular where the orbits degenerate, but if the real parameter is
appropriately restricted to rational values, the metric at the
degeneration surfaces extends smoothly onto a complete and
non-singular compact manifold.  The resulting Einstein-Sasaki spaces
were denoted by $Y^{p,q}$ in \cite{gamaspwa1}, where $p$ and $q$ are
coprime integers with $q<p$.  Further generalisations were obtained in
\cite{gamaspwa3,chlupopo}

    It was shown in \cite{hasaya} that the Einstein-Sasaki spaces
$Y^{p,q}$, and the higher-dimensional generalisations obtained in
\cite{gamaspwa2}, could be obtained by taking certain limits of the
Euclideanised Kerr-de Sitter rotating black hole metrics found in five
dimensions in \cite{hawhuntay}, and in all higher dimensions in
\cite{gilupapo1,gilupapo2}.  Specifically, the limit considered in
\cite{hasaya} involved setting the $n$ independent rotation parameters
of the general $(2n+1)$-dimensional Kerr-de Sitter metrics equal, and
then sending this parameter to a limiting value that corresponds, in
the Lorentzian regime, to having rotation at the speed of light at
infinity.  [This BPS scaling limit for the black hole metrics with two
equal angular momenta was recently studied in \cite{cvgasi}.]

   In a recent paper \cite{cvlupapo}, we showed that vastly greater
classes of complete and non-singular Einstein-Sasaki spaces could be
obtained by starting from the general Euclideanised Kerr-de Sitter
metrics, with unequal rotation parameters, and again taking an
appropriate limit under which the metrics become locally
Einstein-Sasakian.  In fact, this limit can be understood as a
Euclidean analogue of the BPS condition that leads to supersymmetric
black hole metrics.  In five dimensions, this construction leads to
local Einstein-Sasaki metrics of cohomogeneity 2, with $U(1)\times
U(1)\times U(1)$ principal orbits, and two non-trivial real
parameters. In dimension $D=2n+1$, the local Einstein-Sasaki metrics
have cohomogeneity $n$, with $U(1)^{n+1}$ principal orbits, and they
are characterised by $n$ non-trivial real parameters.  In general the
metrics are singular, but by studying the behaviour of the collapsing
orbits at endpoints of the ranges of the inhomogeneous coordinates, we
showed in \cite{cvlupapo} that the metrics extend smoothly onto
complete and non-singular compact manifolds if the real parameters are
appropriately restricted to be rational.  This led to new classes of
Einstein-Sasaki spaces, denoted by $L^{p,q,r_1,\cdots ,r_{n-1}}$ in
$2n+1$ dimensions, where $(p,q,r_1,\ldots,r_{n-1})$ are $(n+1)$
coprime integers.  If the integers are specialised appropriately, the
rotation parameters become equal and the spaces reduce to those
obtained previously in \cite{gamaspwa1} and \cite{gamaspwa2}.  For
example, in five dimensions our general class of Einstein-Sasaki
spaces $L^{p,q,r}$ reduce to those in \cite{gamaspwa1} if $p+q=2r$,
with $Y^{p,q}= L^{p-q,p+q,p}$ \cite{cvlupapo}.

   In this paper, we elaborate on some of our results that appeared in
\cite{cvlupapo}, and give further details about the Einstein-Sasaki
spaces that result from taking BPS limits of the Euclideanised Kerr-de
Sitter metrics.  We also give details about new complete and
non-singular compact Einstein spaces that are not Sasakian, which we
also discussed briefly in \cite{cvlupapo}.  These again arise by
making special choices of the non-trivial parameters in the
Euclideanised Kerr-de Sitter metrics, but this time without first
having taken a BPS limit.  The five-dimensional Einstein-Sasaki spaces
are discussed in section 2, and the higher-dimensional Einstein-Sasaki
spaces in section 3.  In section 4 we discuss the non-Sasakian
Einstein spaces, and the paper ends with conclusions in section 5.  In
an appendix, we discuss certain singular BPS limits, where not all the
rotation parameters are taken to limiting values.  This discussion
also encompasses the case of even-dimensional Kerr-de Sitter metrics,
which do not give rise to non-singular spaces with Killing spinors.

\section{Five-Dimensional Einstein-Sasaki Spaces}\label{es5sec}

\subsection{The local five-dimensional metrics}

   Our starting point is the five-dimensional Kerr-AdS metric found
in \cite{hawhuntay}, which is given by
\bea
ds_5^2 &=& -\fft{\Delta}{\rho^2}\, \Big[ dt -
 \fft{a\, \sin^2\theta}{\Xi_a}\, d\phi - \fft{b\, \cos^2\theta}{\Xi_b}\,
d\psi\Big]^2 + \fft{\Delta_\theta\, \sin^2\theta}{\rho^2}\,
\Big[ a\, dt -\fft{r^2+a^2}{\Xi_a}\, d\phi\Big]^2\nn\\
&& +  \fft{\Delta_\theta\, \cos^2\theta}{\rho^2}\,
\Big[ b\, dt -\fft{r^2+b^2}{\Xi_b}\, d\psi\Big]^2 +
\fft{\rho^2\, dr^2}{\Delta} + \fft{\rho^2\, d\theta^2}{\Delta_\theta}
\nn\\
&& + \fft{(1+ g^2 r^2)}{r^2\, \rho^2}\,
\Big[ a\, b\, dt - \fft{b\, (r^2+a^2)\, \sin^2\theta}{\Xi_a}\, d\phi
- \fft{a\, (r^2+b^2)\, \cos^2\theta}{\Xi_b}\, d\psi\Big]^2\,,
\label{hawkmet}
\eea
where
\bea
\Delta &\equiv & \fft1{r^2}\, (r^2+a^2)(r^2+b^2)(1 + g^2 r^2) -2m\,,
\nn\\
\Delta_\theta &\equiv& 1 - g^2 a^2\, \cos^2\theta -
           g^2 b^2\, \sin^2\theta\,,\nn\\
\rho^2 &\equiv& r^2 + a^2\, \cos^2\theta + b^2\, \sin^2\theta\,,\nn\\
\Xi_a &\equiv& 1 - g^2 a^2\,,\qquad \Xi_b \equiv 1- g^2 b^2\,.
\eea
The metric satisfies $R_{\mu\nu}=-4 g^2\, g_{\mu\nu}$.  As shown in
\cite{gibperpop}, the energy and angular momenta are given by
\be
E =
\fft{\pi\, m\,  (2\Xi_a + 2\Xi_b -\Xi_a\, \Xi_b)}{4 \Xi_a^2 \, \X_b^2}\,,
\qquad
J_a = \fft{\pi\, m\,  a}{2 \Xi_a^2\, \Xi_b}\,,\qquad
 J_b = \fft{\pi\, m\,  b}{2 \Xi_b^2\, \Xi_a}\,.\label{ejrels}
\ee

   As discussed in \cite{cvgilupo}, the BPS limit can be found by
studying the eigenvalues of the \bog matrix arising in the AdS
superalgebra from the anticommutator of the supercharges.  In $D=5$,
these eigenvalues are then proportional to
\be
E \pm g J_a \pm g J_b\,.\label{bog}
\ee
A BPS limit is achieved when one or more of the eigenvalues vanishes.
For just one zero eigenvalue, the four cases in (\ref{bog}) are
equivalent under reversals of the angular velocities, so we may
without loss of generality consider $E-g J_a -g J_b=0$.  From
(\ref{ejrels}), we see that this is achieved by taking a limit in
which $g a$ and $g b$ tend to unity, namely, by setting $g
a=1-\ft12\epsilon \alpha$, $gb=1-\ft12\epsilon\beta$, rescaling $m$
according to $m=m_0\epsilon^3$, and sending $\epsilon$ to zero.  As we
shall see, the metric remains non-trivial in this limit.  An
equivalent discussion in the Euclidean regime leads to the conclusion
that in the corresponding limit, one obtains five-dimensional Einstein
metrics admitting a Killing spinor.

   We perform a Euclideanisation of (\ref{hawkmet}) by making the
analytic continuations
\be
t\rightarrow \im t\,,\quad g\rightarrow \fft{\im }{\sqrt{\lambda}}\,,
\quad a\rightarrow \im a\,,\quad b\rightarrow \im b\,,
\ee
and then take the BPS limit by setting
\bea
&&a=\lambda^{-\ft12} (1 - \ft12\alpha\,\epsilon)\,,\quad
b=\lambda^{-\ft12} (1 - \ft12\beta\,\epsilon)\,,\nn\\
&&r^2=\lambda^{-1} (1 - x\epsilon)\,,\quad
m=\ft12\lambda^{-1} \mu \epsilon^3
\eea
and sending $\epsilon\rightarrow 0$.  The metric becomes
\be
\lambda\,ds_5^2 = (d\tau + \sigma)^2 + ds_4^2\,,\label{5met}
\ee
where
\bea
ds_4^2 &=& \fft{\rho^2\,dx^2}{4\Delta_x} +
\fft{\rho^2\,d\theta^2}{\Delta_\theta} +
\fft{\Delta_x}{\rho^2} (\fft{\sin^2\theta}{\alpha} d\phi +
\fft{\cos^2\theta}{\beta} d\psi)^2\nn\\
&& + \fft{\Delta_\theta\sin^2\theta\cos^2\theta}{\rho^2}
  (\fft{\alpha - x}{\alpha}
d\phi - \fft{\beta - x}{\beta} d\psi)^2\,,\nn\\
\sigma &=& \fft{(\alpha -x)\sin^2\theta}{\alpha} d\phi +
\fft{(\beta-x)\cos^2\theta}{\beta} d\psi\,,\label{d4met}\\
\Delta_x &=& x (\alpha -x) (\beta - x) - \mu\,,\quad
    \rho^2=\Delta_\theta-x\,,\nn\\
\Delta_\theta &=& \alpha\, \cos^2\theta + \beta\, \sin^2\theta\,.
\nn
\eea
A straightforward calculation shows that the four-dimensional metric
in (\ref{d4met}) is Einstein.  Note that the parameter $\mu$ is
trivial, and can be set to any non-zero constant, say, $\mu=1$, by
rescaling $\alpha$, $\beta$ and $x$.  The metrics depend on two
non-trivial parameters, which we can take to be $\alpha$ and $\beta$
at fixed $\mu$.  However, it is sometimes convenient to retain $\mu$,
allowing it to be determined as the product of the three roots $x_i$
of $\Delta_x$.

   It is also straightforward to verify that the four-dimensional
Einstein metric in (\ref{d4met}) is K\"ahler, with K\"ahler form
$J=\ft12 d\sigma$.  We find that
\be
J = e^1\wedge e^2 + e^3\wedge e^4\,,
\ee
when expressed in terms of the vielbein
\bea
e^1 &=& \fft{\rho dx}{2\sqrt{\Delta_x}}\,,\qquad
e^2 = \fft{\sqrt\Delta_x}{\rho}\, (\fft{\sin^2\theta}{\alpha}\, d\phi+
    \fft{\cos^2\theta}{\beta}\, d\psi)\,,\nn\\
e^3 &=& \fft{\rho d\theta}{\sqrt{\Delta_\theta}}\,,\qquad
e^4 = \fft{\sqrt{\Delta_\theta}\, \sin\theta\cos\theta}{\rho}\, 
  (\fft{\alpha - x}{\alpha} d\phi - \fft{\beta - x}{\beta} d\psi)\,.
\eea
A straightforward calculation confirms that $J$ is indeed covariantly 
constant.

\subsection{Global structure of the five-dimensional solutions}

   Having obtained the local form of the five-dimensional
Einstein-Sasaki metrics, we can now turn to an analysis of the global
structure.  The metrics are in general of cohomogeneity 2, with toric
principal orbits $U(1)\times U(1)\times U(1)$.  The orbits degenerate
at $\theta=0$ and $\theta=\ft12 \pi$, and at the roots of the cubic
function $\Delta_x$ appearing in (\ref{d4met}).  In order to obtain
metrics on complete non-singular manifolds, one must impose
appropriate conditions to ensure that the collapsing orbits extend
smoothly, without conical singularities, onto the degenerate surfaces.
If this is achieved, one can obtain a metric on a non-singular
manifold, with $0\le\theta\le\ft12\pi$ and $x_1\le x\le x_2$, where
$x_1$ and $x_2$ are two adjacent real roots of $\Delta_x$. In fact,
since $\Delta_x$ is negative at large negative $x$ and positive at
large positive $x$, and since we must also have $\Delta_x>0$ in the
interval $x_1<x<x_2$, it follows that $x_1$ and $x_2$ must be the
smallest two roots of $\Delta_x$.

   The easiest way to analyse the behaviour at each collapsing orbit
is to examine the associated Killing vector $\ell$ whose length
vanishes at the degeneration surface. By normalising the Killing
vector so that its ``surface gravity'' $\kappa$ is equal to unity, one
obtains a translation generator $\del/\del \chi$ where $\chi$ is a
local coordinate near the degeneration surface, and the metric extends
smoothly onto the surface if $\chi$ has period $2\pi$.  The ``surface
gravity'' for the Killing vector $\ell$ is given, in the Euclidean
regime, by
\be
\kappa^2 = -\fft{g^{\mu\nu}\, (\del_\mu \ell^2)(\del_\nu
\ell^2)}{4\ell^2}
\ee
in the limit that the degeneration surface is reached.

   The normalised Killing vectors that vanish at the degeneration
surfaces $\theta=0$ and $\theta=\ft12\pi$ are simply given by
$\del/\del \phi$ and $\del/\del\psi$ respectively.  At the
degeneration surfaces $x=x_1$ and $x=x_2$, we find that the associated
normalised Killing vectors $\ell_1$ and $\ell_2$ are given by
\be
\ell_i = c_i\, \fft{\del}{\del \tau} + a_i\, \fft{\del}{\del\phi} +
            b_i\, \fft{\del}{\del\psi}\,,\label{ells}
\ee
where the constants $c_i$, $a_i$ and $b_i$ are given by
\bea
a_i &=& \fft{\alpha c_i}{x_i - \alpha}\,,\qquad
b_i = \fft{\beta c_i}{x_i-\beta}\,,\nn\\
c_i &=& \fft{(\alpha-x_i)(\beta-x_i)}{2(\alpha+\beta) x_i -
\alpha\beta - 3 x_i^2}\,.\label{abci}
\eea

   Since we have a total of four Killing vectors $\del/\del\phi$,
$\del/\del\psi$, $\ell_1$ and $\ell_2$ that span a three-dimensional
space, there must exist a linear relation amongst them.  Since they
all generate translations with a $2\pi$ period repeat, it follows that
unless the coefficients in the linear relation are rationally related,
then by taking integer combinations of translations around the $2\pi$
circles, one could generate a translation implying an identification
of arbitrarily nearby points in the manifold.  Thus one has the
requirement for obtaining a non-singular manifold that the linear
relation between the four Killing vectors must be expressible as
\be
p \ell_1 + q \ell_2 + r\, \fft{\del}{\del\phi} + 
s \, \fft{\del}{\del\psi}=0\label{lincomb}
\ee
for {\it integer} coefficients $(p,q,r,s)$, which may, of course, be
assumed to be coprime.  We must also require that all subsets of three
of the four integers be coprime too.  This is because if any three had
a common divisor $k$, then dividing (\ref{lincomb}) by $k$ one could
deduce that the direction associated with the Killing vector whose
coefficient was not divisible by $k$ would be identified with period
$2\pi/k$, thus leading to a conical singularity.

   From (\ref{lincomb}), and (\ref{ells}), we have
\bea
&& p a_1 + q a_2 + r=0\,,\qquad
p b_1 + q b_2 + s=0\,,\nn\\
&& p c_1 + q c_2=0\,.\label{klmn}
\eea
From these relations it then follows that the ratios
between each pair of the four quantities
\be
a_1 c_2-a_2 c_1\,,\quad
 b_1 c_2 -b_2 c_1 \,,\quad
c_1\,,\quad c_2\label{rationals}
\ee
must be rational.  Thus in order to obtain a metric that extends
smoothly onto a complete and non-singular manifold, we must choose the
parameters in (\ref{d4met}) so that the rationality of the ratios is
achieved.  In fact it follows from (\ref{abci}) that
\be
1+a_i + b_i + 3 c_i = 0\label{abcid}
\ee
for all roots $x_i$, and using this one can show that there are only
two independent rationality conditions following from the requirements
of rational ratios for the four quantities in (\ref{rationals}).  One
can also see from (\ref{abcid}) that
\be
p+q-r-s=0\,,\label{klmnrel}
\ee
and so the further requirement that all triples chosen from the
coprime integers $(p,q,r,s)$ also be coprime is automatically
satisfied.

   The upshot from the above discussion is that we can have complete
and non-singular five-dimensional Einstein-Sasaki spaces $L^{p,q,r}$,
where
\be
pc_1+q c_2=0\,,\quad pa_1 + q a_2  + r=0\,.
\ee
These equations and (\ref{abcid}) allow one to solve for $\alpha$,
$\beta$ and the roots $x_1$ and $x_2$, for positive coprime integer
triples $(p,q,r)$.  The requirements $0<x_1\le x_2<x_3$, and
$\alpha>x_2$, $\beta>x_2$, restrict the integers to the domain $0< p
\le q$ and $0 <r < p+q$.  All such coprime triples yield complete and
non-singular Einstein-Sasaki spaces $L^{p,q,r}$, and so we get
infinitely many new examples.

   The spaces $L^{p,q,r}$ all have the topology of $S^2\times S^3$.
We are very grateful to Krzysztof Galicki for the following argument
which shows this: The total space of the Calabi-Yau cone, with metric
$ds_6^2= dy^2 + y^2\, \lambda ds_5^2$, can be viewed as a circle
reduction (\ie a symplectic quotient) of $\C^4$ by the diagonal action
of $S^1(p,q,-r,-s)$ with $p+q-r-s=0$.  The topology of the $L^{p,q,r}$
spaces has also been discussed in detail in \cite{martspar}.

   The volume of $L^{p,q,r}$ (with $\lambda=1$) is given by
\be
V=\fft{\pi^2(x_2-x_1)
(\alpha+\beta -x_1-x_2)\Delta\tau}{2k\alpha\beta}\,,
\ee
where $\Delta\tau$ is the period of the coordinate $\tau$, and
$k=\hbox{gcd}\, (p,q)$. Note that the $(\phi,\psi)$ torus is factored
by a freely-acting $\Z_k$, along the diagonal.  $\Delta\tau$ is given
by the minimum repeat distance of $2\pi c_1$ and $2\pi c_2$, \ie the
minimisation of $|2\pi c_1 M + 2\pi c_2 N|$ over the integers $(M,N)$.
We have
\be
2\pi(c_1 M+ c_2 N) = \fft{2\pi c_1}{q}\, (M q - N p)\,,
\ee
and so if $p$ and $q$ have greatest common divisor $k$, the integers
$M$ and $N$ can be chosen so that the minimum of $|M q - N p|$ is $k$,
and so
\be
\Delta\tau=\fft{2\pi k |c_1|}{q}\,.
\ee
The volume of $L^{p,q,r}$ is therefore given by
\be
V=\fft{\pi^3 |c_1| (x_2-x_1)
(\alpha+\beta -x_1-x_2)}{q\alpha\beta}\,,
\ee

   There is a quartic equation expressing
$V$ purely in terms of $(p,q,r)$.  Writing 
\be
V=\fft{\pi^3 (p+q)^3 W}{8pqrs}\,, \label{VWrel}
\ee
we find
\bea
0&=& 27 W^4-8(2-9h_+) W^3- [8h_+(2-h_+)^2-h_-^2(30+9h_+)]W^2\nn\\
&& - 2h_-^2[2(2-h_+)^2-3h_-^2]W -(1-f^2)(1-g^2)h_-^4\label{quartic}
\eea
where 
\be
f\equiv \fft{q-p}{p+q}\,,\qquad g \equiv\fft{r-s}{p+q}\,,\label{fgdef}
\ee
and $h_\pm=f^2\pm g^2$.  The
central charge of the dual field theory is rational if $W$ is rational,
which, as we shall show below, is easily achieved.

   If one sets $p+q=2r$, implying that $\alpha$ and $\beta$ become
equal, our Einstein-Sasaki metrics reduce to those in
\cite{gamaspwa1}, and the conditions we have discussed for achieving
complete non-singular manifolds reduce to the conditions for the
$Y^{p,q}$ obtained there, with $Y^{p,q}= L^{p-q,p+q,p}$.  The quartic
(\ref{quartic}) then factorises over the rationals into quadrics,
giving the volumes found in \cite{gamaspwa1}.

   Further special limits also arise.  For example, if we take
$p=q=r=1$, the roots $x_1$ and $x_2$ coalesce, $\alpha=\beta$, and the
metric becomes the homogeneous $T^{1,1}$ space, with the
four-dimensional base space being $S^2\times S^2$.  In another limit,
we can set $\mu=0$ in (\ref{d4met}) and obtain the round metric on
$S^5$, with $CP^2$ as the base. (In fact, we obtain $S^5/Z_q$ if
$p=0$.) Except in these special ``regular'' cases, the
four-dimensional base spaces themselves are singular, even though the
Einstein-Sasaki spaces $L^{p,q,r}$ are non-singular.  The
Einstein-Sasaki space is called quasi-regular if $\del/\del\tau$ has
closed orbits, which happens if $c_1$ is rational.  If $c_1$ is
irrational the orbits of $\del/\del\tau$ never close, and the
Einstein-Sasaki space is called irregular.

\subsection{Quasi-regular examples}\label{qregsec}

   We find that we can obtain quasi-regular Einstein-Sasaki 5-spaces,
with rational values for the roots $x_i$, the parameters $\alpha$,
$\beta$, and the volume factor $W$ if the integers $(p,q,r)$ are
chosen such that
\be
\fft{q-p}{p+q}= \fft{2(v-u)(1+ u v)}{4- (1+u^2)(1+v^2)}\,,\qquad
\fft{r-s}{p+q}= \fft{2(v+u)(1- u v)}{4- (1+u^2)(1+v^2)}\,,\label{uv1}
\ee
where $u$ and $v$ are any rational numbers  satisfying
\be
 0<v<1\,,\qquad -v <u < v\,.\label{uvregion}
\ee
A convenient choice that eliminates the redundancy in the
$(\alpha,\beta,\mu)$ parameterisation of the local solutions is by
taking $x_3=1$, in which case we then have rational solutions with
\bea
x_1&=& \ft14(1+u)(1-v)\,,\qquad
x_2= \ft14 (1-u)(1+v)\,,\qquad
x_3= 1\,,\label{rootsuv}\\
\alpha&=&1 -\ft14 (1+u)(1+v)\,,\qquad \beta =1-\ft14(1-u)(1-v)\,,\qquad
    \mu= \ft1{16} (1-u^2)(1-v^2)\,.\nn
\eea
 From these, we have
\bea
c_1 &=& -\fft{2(1-u)(1+v)}{(v-u)[4-(1+u)(1-v)]}\,,\qquad 
c_2= \fft{2(1+u)(1-v)}{(v-u)[4-(1-u)(1+v)]}\,,\nn\\
a_1&=&\fft{(1+v)(3-u-v-uv)}{(v-u)[4-(1+u)(1-v)]}\,,\qquad
a_2= - \fft{(1+u)(3-u-v-uv)}{(v-u)(4-(1-u)(1+v)]}\,,\nn\\
b_1 &=& \fft{(1-u)(3 +u + v-uv)}{(v-u)[4-(1+u)(1-v)]}\,,\qquad
b_2=- \fft{(1-v)(3+u+v-uv)}{(v-u)[4-(1-u)(1+v)]}\,.
\eea
It follows that $c_1$ is also rational, and so these Einstein-Sasaki
spaces are quasi-regular, with closed orbits for $\del/\del\tau$.  The
volume is given by (\ref{VWrel}), with
\be
W= \fft{16(1-u^2)^2\, (1-v^2)^2}{(3- u^2 - v^2 - u^2 v^2)^3}\,,
\ee
and so the ratio of $V$ to the volume of the unit 5-sphere (which is
$\pi^3$) is rational too.

   Note that although we introduced the $(u,v)$ parameterisation in
order to write quasi-regular examples with rational roots and volumes,
the same parameterisation is also often useful in general.  One simply
takes $u$ and $v$ to be real numbers, not in general rational, defined
in terms of $p$, $q$, $r$ and $s$ by (\ref{uv1}).  They are again
subject to the restrictions (\ref{uvregion}).

\subsection{Volumes and the Bishop bound}

   Note that the volume $V$ can be expressed in terms of $u$, $v$ and
$p$ as
\be
V = \fft{16\pi^3 (1+u)(1-v)}{p\, (3+u+v-uv)(3+u-v+uv)(3-u-v-uv)}\,,
\ee
where $u$ and $v$ are given in terms of $p$, $q$ and $r$ by \ref{uv1}.
It is easy to verify that the volume is always bounded above by the
volume of the unit 5-sphere,\footnote{Recall that all our volume
formulae are with respect to spaces normalised to $R_{ij} = 4
g_{ij}$.} as it must be by Bishop's theorem \cite{thebish}.  To see
this, define
\be
Y\equiv 1- \fft{p V}{\pi^3} \,.
\ee
Since $p$ is a positive integer, then if we can show that $Y>0$ for
all our inhomogeneous Einstein-Sasaki spaces, it follows that they
must all have volumes less than $\pi^3$, the volume of the unit
$S^5$. It is easy to see that
\be
Y = \fft{(1-u)(1+v) \, F }{ (3+u+v-uv)(3+u-v+uv)(3-u-v-uv)}\,,
\ee
where
\be
F = 11 + (u+v)^2 + 2uv + 4(u-v) - u^2 v^2\,.
\ee
With $u$ and $v$ restricted to the region defined by (\ref{uvregion}), 
it is clear that the sign of $Y$ is the same as the sign of $F$.  It 
also follows from (\ref{uvregion}) that
\be
2uv > -2\,,\qquad 4(u-v) > -8\,,\qquad - u^2 v^2 > -1\,,
\ee
and so we have $F>0$.  Thus $Y >0$ for all the inhomogeneous
Einstein-Sasaki spaces, proving that they all satisfy $V<\pi^3$.

\subsection{The $u\leftrightarrow -u$ symmetry}

   By making appropriate redefinitions of the coordinates, we can make
manifest the discrete symmetry of the five-dimensional metrics under
the transformation $u\rightarrow -u$, which, from (\ref{uv1}),
corresponds to the exchange of the integers $(p,q)$ with the integers
$(r,s)$.  Accordingly, we define new coordinates\footnote{A similar
redefinition of coordinates has also been given in \cite{martspar}.}
\be
y= \Delta_\theta\,,\qquad \hat\psi= \fft{\phi-\psi}{\beta-\alpha}\,,\qquad
\hat \phi = \fft{\alpha^{-1}\, \phi-\beta^{-1} \psi}{\beta-\alpha}\,,\qquad
\hat\tau = \tau + \fft{\beta\psi-\alpha \phi}{\beta-\alpha}\,.
\ee
In terms of these, the Einstein-Sasaki metrics become
\be
\lambda ds_5^2 = (d\hat \tau + \hat\sigma)^2 + ds_4^2\,,\label{5metxy}
\ee
with
\bea
ds_4^2 &=& \fft{(y-x)dx^2}{4 \Delta_x} + \fft{(y-x) dy^2}{4\Delta_y}
 + \fft{\Delta_x}{y-x}\, (d\hat\psi- y d\hat\phi)^2 + 
   \fft{\Delta_y}{y-x}\, (d\hat\psi - x d\hat\phi)^2\,,\nn\\
\hat\sigma &=& (x+y) d\hat\psi - xy d\hat\phi\,.
\eea
The metric functions $\Delta_x$ and $\Delta_y$ are given by
\be
\Delta_x = x(\alpha-x)(\beta-x) -\mu\,,\qquad
\Delta_y = -y(\alpha-y)(\beta-y)\,,
\ee

   It is convenient now to adopt the parameterisation introduced in 
(\ref{uv1}).  Note that this can be done whether or not $u$ and $v$ 
are chosen to be rational.  Then from (\ref{rootsuv}) we have
\bea
\Delta_x &=& \ft1{16}(x-1)[4x- (1-u)(1+v)][4x-(1+u)(1-v)]\,,\nn\\
\Delta_y &=& -\ft1{16} y[4(1-y)- (1-u)(1-v)][4(1-y)-(1+u)(1+v)]\,.
\eea
It is now manifest that the five-dimensional metric (\ref{5metxy}) is
invariant under sending $u\rightarrow -u$, provided that at the same
time we make the coordinate transformations
\bea
x &\longrightarrow& 1-y\,,\qquad y \longrightarrow 1-x\,,\nn\\
\hat \psi &\longrightarrow& -\hat\psi + \hat\phi\,,\qquad
   \hat\tau \longrightarrow \hat\tau + 2\hat\psi -\hat\phi\,.
\eea

   The argument above shows that to avoid double counting, we should
further restrict the coprime integers $(p,q,r)$ so that either
$u\ge0$, or else $u\le 0$.  We shall make the latter choice, which
implies that we should restrict $(p,q,r)$ so that
\be
0\le p\le q\le r \le p+q\,.
\ee

\subsection{The $u=0$ case}

     Having seen that there is a symmetry under $u\leftrightarrow -u$,
it is of interest to study the fixed point of this discrete symmetry,
\ie $u=0$, which corresponds to setting $q=r$ (and hence $p=s$).

   From (\ref{rootsuv}), a convenient way of restricting to $u=0$ is
by choosing 
\be
\mu= \ft2{27} (2\alpha-\beta)(2\beta-\alpha)(\alpha+\beta)\,.
\ee
This allows us to factorise the function $\Delta_x$, giving
\be
\Delta_x = \ft1{27} (2\alpha-\beta-3x)(\alpha-2\beta+3x)
      (2\alpha+2\beta -3x)\,.
\ee

   Now we introduce new tilded coordinates, defined by
\bea
\td x &=& -x + \ft13(\alpha+\beta)\,,\qquad 
\td y= \Delta_\theta - \ft13 (\alpha+\beta)\,,\nn\\
\phi &=& -\alpha[\td\psi + \ft13(2\beta-\alpha) \td\phi]\,,\qquad
\psi = - \beta[ \td\psi + \ft13(2\alpha-\beta)\td\phi]\,,\nn\\
\td\tau &=& \tau -\ft13(\alpha+\beta)\td\psi + 
         \ft19 (2\alpha^2 -5\alpha\beta + 2\beta^2)\td\phi\,.
\eea
After doing this, the metric takes on the form
\be
\lambda ds_5^2 = (d\td\tau+\td\sigma)^2 + ds_4^2\,,
\ee
with 
\bea
ds_4^2 &=& \fft{(\td x+\td y)d\td x^2}{4 \Delta(\td x)} + 
       \fft{(\td x+ \td y) d\td y^2}{4\Delta(\td y)}
 + \fft{\Delta(\td x)}{\td x+\td y}\, (d\td\psi+ \td y d\td \phi)^2 + 
   \fft{\Delta(\td y)}{\td x+\td y}\, (d\td\psi - \td x d\td\phi)^2\,,
\nn\\
\td\sigma &=& (\td y-\td x) d\td\psi - \td x \td yd\td\phi\,.
\eea
The metric functions $\Delta(\td x)$ and $\Delta(\td y)$, which are
now the same function with either $\td x$ or $\td y$ as argument, are
given by
\be
\Delta(z) = \ft1{27} (\alpha-2\beta + 3z)(2\alpha-\beta-3z)
                         (\alpha+\beta+3z)\,.
\ee
It is therefore natural to define new parameters $\gamma_1$ and
$\gamma_2$, given by
\be
\gamma_1=\ft13(2\alpha-\beta)\,,\qquad 
\gamma_2= \ft13(2\beta-\alpha)\,.
\ee
In terms of these, the function $\Delta(z)$ becomes
\be
\Delta(z) = -(\gamma_1 -z)(\gamma_2-z)(\gamma_1+\gamma_2+z)\,.
\ee

   There is now a manifest discrete symmetry, under which we send
\be
\td x\leftrightarrow \td y\,,\qquad \td\phi\leftrightarrow -\td\phi\,.
\ee
It is worth remarking that the quartic polynomial (\ref{quartic})
determining the volume factorises in quadrics over the rationals in
the case that $u=0$, giving
\be
V=\fft{4\pi^3}{27p^2q^2} \Big((p+q)(p-2q)(q-2p) +
2 (p^2 -pq + q^2)^{3/2}\Big)\,.
\ee

\subsection{Curvature invariants}

    In this section, we present some results for curvature invariants
for the five-dimensional Einstein-Sasaki metrics.  We find
\bea
I_2 &\equiv & R^{ijk\ell} \, R_{ijk\ell} = 
              \fft{192(\rho^{12} + 2\mu^2)}{\rho^{12}}\,,\nn\\
I_3 &\equiv& R^{ij}{}_{k\ell}\, R^{k\ell}{}_{mn}\, R^{mn}{}_{ij}=
     \fft{384(5\rho^{18} + 12\mu^2\rho^6 + 8 \mu^3)}{\rho^{18}}\,,\nn\\
J_3&\equiv & R^i{}_j{}^k{}_\ell\, R^j{}_m{}^\ell{}_n\, R^m{}_i{}^n{}_k
= \fft{96(\rho^{18} - 12\mu^2 \rho^6 + 16\mu^3)}{\rho^{18}}\,.
\eea

   Since these curvature invariants depend on the coordinates only via
the single combination $\rho^2= \alpha\sin^2\theta + \beta\cos^2\theta
-x$, one might wonder whether the Einstein-Sasaki metrics, despite
ostensibly being of cohomogeneity 2, were actually only of
cohomogeneity 1, becoming manifestly so when described in an
appropriate coordinate system.  In fact this is not the case, as can
be seen by calculating the scalar invariant
\be
K = g^{\mu\nu} (\del_\mu I_2) (\del_\nu I_2)\,,
\ee
which turns out to be given by
\bea
K &=& \fft{2^{18} 3^4 \mu^4}{\rho^{30}}\Big[-\rho^6 +
(2\beta-\alpha) \rho^4 -\beta(\beta-\alpha)\rho^2  -\mu\nn\\
&& \qquad\qquad+ 
  (\alpha-\beta)[3\rho^2 - 2(2\beta-\alpha) ]\rho^2\, \cos^2\theta
        -3(\alpha-\beta)^2\, \rho^2\, \cos^4\theta\Big]\,.
\eea
Since this invariant does not depend on the $x$ and $\theta$
coordinates purely via the combination $\rho^2= \alpha\sin^2\theta +
\beta\cos^2\theta -x$, we see that the metrics do indeed genuinely
have cohomogeneity 2.  They do, of course, reduce to cohomogeneity 1
if the parameters $\alpha$ and $\beta$ are set equal.

\section{Higher-Dimensional Einstein-Sasaki Spaces}\label{esnsec}

   The construction of five-dimensional Einstein-Sasaki spaces that we
have given in section \ref{es5sec} can be extended straightforwardly
to all higher odd dimensions.  We take the rotating Kerr-de Sitter
metrics obtained in \cite{gilupapo1,gilupapo2}, and impose the
Bogomol'nyi conditions $E- g\sum_i J_i=0$, where $E$ and $J_i$ are the
energy and angular momenta that were calculated in \cite{gibperpop},
and given in (\ref{ejrels}).  We find that a non-trivial BPS limit
exists where $g a_i = 1 - \ft12\alpha_i \epsilon$, $m= m_0
\epsilon^{n+1}$.  After Euclideanisation of the $D=2n+1$ dimensional
rotating black hole metrics obtained in \cite{gilupapo1}, which is
achieved by sending $\tau\rightarrow \im\, \tau$, and $a_i\rightarrow
\im\, a_i$ in equation (3.1) of that paper (and using $y$ rather than
$r$ as the radial variable, to avoid a clash of notations later), one
has
\bea
ds^2 &=&  W\, (1 -\lambda\,y^2)\,
d\tau^2 + \fft{U\, dy^2}{V-2m} +
\fft{2m}{U}\Bigl(d\tau - \sum_{i=1}^n \fft{a_i\, \mu_i^2\, d\varphi_i}{
1 - \lambda\, a_i^2}\Bigr)^2 \label{blodd}\\
&&
+ \sum_{i=1}^n \fft{y^2 - a_i^2}{1 - \lambda\, a_i^2}
\, [d\mu_i^2 + \mu_i^2\, (d\varphi_i +\lambda\, a_i\, d\tau)^2] +
\fft{\lambda}{W\, (1-\lambda y^2)}
\Big( \sum_{i=1}^n \fft{(y^2 - a_i^2)\mu_i\, d\mu_i}{
1 - \lambda\, a_i^2}\Big)^2 \,,\nn
\eea
where
\bea
V &\equiv& \fft1{y^2}\, (1-\lambda\, y^2)\, \prod_{i=1}^n (y^2 - a_i^2)
\,,\qquad
W \equiv \sum_{i=1}^n \fft{\mu_i^2}{1-\lambda\, a_i^2}\,,\nn\\
U &=& \sum_{i=1}^n \fft{\mu_i^2}{y^2 - a_i^2}\,
\prod_{j=1}^n (y^2 - a_j^2)\,.
\eea

   The BPS limit is now achieved by setting
\bea
&& a_i=\lambda^{-\ft12} (1 - \ft12\alpha_i\,\epsilon)\,,\nn\\
&&y^2=\lambda^{-1} (1 - x\epsilon)\,,\quad
m=\ft12\lambda^{-1} \mu \epsilon^{n+1}\,,\label{dbpslim}
\eea
and then sending $\epsilon\rightarrow 0$.  We then obtain $D=2n+1$ 
dimensional Einstein-Sasaki
metrics $ds^2$, given by
\be
\lambda ds^2 = (d\tau+\sigma)^2 + d\bar s^2\,,\label{esgenmet}
\ee
with $R_{\mu\nu}=2n\lambda g_{\mu\nu}$, where the $2n$-dimensional
metric $d\bar s^2$ is Einstein-K\"ahler, with K\"ahler form
$J=\ft12d\sigma$, and
\bea
d\bar s^2 &=& \fft{Y dx^2}{4x F} - \fft{x(1-F)}{Y}
\Big(\sum_i \alpha_i^{-1}\,  \mu_i^2 d\varphi_i\Big)^2 
 + \sum_i (1-\alpha_i^{-1}\, x)(d\mu_i^2 + \mu_i^2 d\varphi_i^2)
\nn\\
&&+ \fft{x}{\sum_i \alpha_i^{-1} \mu_i^2}\,
        \Big( \sum_j \alpha_j^{-1}\, \mu_j d\mu_j\Big)^2
-\sigma^2\,,\nn\\
\sigma &=& \sum_i (1-\alpha_i^{-1}x)\mu_i^2\, d\varphi_i\,,\\
Y&=&\sum_i\fft{\mu_i^2}{\alpha_i-x}\,,\qquad
  F= 1- \fft{\mu}{x}\, \prod_i(\alpha_i-x)^{-1}\,,\nn
\eea
where $\sum_i \mu_i^2=1$.  The $D=2n+1$ dimensional Einstein-Sasaki
metrics have cohomogeneity $n$, with $U(1)^{n+1}$ principal orbits.

    The discussion of the global properties is completely analogous to
the one we gave previously for the five-dimensional case.  The $n$
Killing vectors $\del/\del\varphi_i$ vanish at the degenerations of
the $U(1)^{n+1}$ principal orbits where each $\mu_i$ vanishes, and
conical singularities are avoided if each coordinate $\varphi_i$ has
period $2\pi$.  The Killing vectors
\be
\ell_i = c(i) \,\fft{\del}{\del\tau} +
\sum_j b_j(i)\, \fft{\del}{\del\varphi_j}
\ee
vanish at the roots $x=x_i$ of $F(x)$, and have unit surface gravities
there, where
\be
b_j(i) = -\fft{c(i) \alpha_j}{\alpha_j-x_i}\,,\quad
c(i)^{-1} = \sum_j \fft{x_i}{\alpha_j-x_i} -1\,.\label{acdefs}
\ee
The metrics extend smoothly onto complete and non-singular manifolds
if 
\be
p \ell_1 + q\ell_2 + \sum_j r_j \fft{\del}{\del\varphi_j}=0
\ee
for coprime integers $(p,q,r_j)$, where in addition all possible
subsets of $(n+1)$ of the integers are also coprime (which is again
automatic--see below).  This implies the algebraic equations
\be
p \, c(1) + q \, c(2)=0\,,\quad
p \, b_j(1) + q \, b_j(2) + r_j=0\,,\label{bcpqrj}
\ee
determining the roots $x_1$ and $x_2$, and the parameters $\alpha_j$.
The two roots of $F(x)$ must be chosen so that $F>0$ when $x_1<x<x_2$.
With these conditions satisfied, we obtain infinitely many new
complete and non-singular compact Einstein-Sasaki spaces in all odd
dimensions $D=2n+1$.

   It follows from (\ref{acdefs}) that 
\be
\sum_j b_j(i) + (n+1) c(i) + 1=0\,,
\ee
and hence using (\ref{bcpqrj}) we have
\be
p+q=\sum_j r_j\,. \label{pqrj}
\ee
This can be used to eliminate $r_n$ in favour of the other $(n+1)$
integers.  The Einstein-Sasaki spaces, which we denote by
$L^{p,q,r_1,\cdots ,r_{n-1}}$, are therefore characterised by
specifying $(n+1)$ coprime integers, which must lie in an appropriate
domain.  Without loss of generality, we may choose $p<q$, and order
the two roots $x_1$ and $x_2$ so that $x_1<x_2$.  It follows that we
shall have
\be
c_1<0\,,\qquad c_2>0\,,\qquad |c_1| > c_2\,.
\ee
The parameters $\alpha_j$ must all satisfy $\alpha_j>x_2$, to ensure
that $Y$ is always positive.  From (\ref{bcpqrj}) we have therefore
have
\be
r_j = \fft{q c(2)\, \alpha_j (x_2-x_1)}{(\alpha_j-x_1)(\alpha_j-x_2)}>0\,.
\ee
To avoid overcounting, we can therefore specify the domain by
\be
0<p<q\,, \qquad 0<r_1 \le r_2 \le \cdots \le  r_{n-1} \le r_n\,.
\ee

   The $n$-torus of the $\varphi_j$ coordinates is in general factored
by a freely-acting $\Z_k$, where $k=\hbox{gcd}\, (p,q)$.  The volume
(with $\lambda=1$) is given by
\be
V= \fft{|c(1)|}{q}\, {\cal A}_{2n+1}\, \Big[\prod_i
 \Big(1-\fft{x_1}{\alpha_i}\Big) -\prod_i\Big(1-\fft{x_2}{\alpha_i}
         \Big)\Big]\,,
\ee
since $\Delta\tau$ is given by $2\pi k|c(1)|/q$, where ${\cal
A}_{2n+1}$ is the volume of the unit $(2n+1)$-sphere.  In the special
case that the rotations $\alpha_i$ are set equal, the metrics reduce
to those obtained in \cite{gamaspwa2}.

\section{Non-Sasakian Einstein Spaces}

   So far in this paper, we have concentrated on the situations, in
odd dimensions, where a limit of the Euclideanised Kerr-de Sitter
metrics can be taken in which one has a Killing spinor.  In this
section, we shall describe the more general situation in which no
limit is taken, and so one has Einstein metrics that do not have
Killing spinors.  They are therefore Einstein spaces that are not
Sasakian.  Again, the question arises as to whether these metrics can,
for suitable choices of the parameters, extend smoothly onto complete
and non-singular compact manifolds.  As in the previous discussion in
the Einstein-Sasaki limit, this question reduces to whether smooth
extensions onto the surfaces where the principal orbits degenerate are
possible.

    This question was partially addressed in
\cite{gilupapo1,gilupapo2}, where the problem was studied in the case
that the two roots defining the endpoints of the range of the radial
variable were taken to be coincident.  This ensured that the surface
gravities at the endpoints of the (rescaled) radial variable were
equal in magnitude.  However, as we have seen in the discussion for
the Einstein-Sasaki limits, the requirement of equal surface gravities
is more restrictive than is necessary for obtaining non-singular
spaces.  In this section, we shall study the problem of obtaining
non-singular spaces within this more general framework.

\subsection{Odd dimensions}

   The Euclideanised Kerr-de Sitter metrics in odd dimensions $D=2n+1$ 
are given in equation (\ref{blodd}). From the results in 
\cite{gilupapo1,gilupapo2}, the Killing vector
\be
\td\ell \equiv \fft{\del}{\del\tau} -\sum_{j=1}^n
 \fft{a_j\, (1-\lambda y_0^2)}{y_0^2 -a_j^2} \,\fft{\del}{\del\varphi_j}
\ee
has vanishing norm at a root $y=y_0$ of $V(y)-2m=0$, and it has a
``surface gravity'' given by
\be
\kappa = y_0\, (1-\lambda y_0^2)\, \sum_{j=1}^n \fft{1}{y_0^2-a_j^2} - 
           \fft1{y_0}\,.
\ee

    Following the strategy we used for studying the degenerate orbits
of the metrics in the Einstein-Sasaki limit, we now introduce a
rescaled Killing vector $\ell = c\, \td\ell$ with $c$ chosen so that
$\ell$ has unit surface gravity. Thus we define Killing vectors
\bea
\ell_1 &=& c(1) \, \fft{\del}{\del\tau} + \sum_{j=1}^n b_j(1)\, 
       \fft{\del}{\del\varphi_j}\,,\nn\\
\ell_2 &=& c(2) \, \fft{\del}{\del\tau} + \sum_{j=1}^n b_j(2)\, 
       \fft{\del}{\del\varphi_j}\,,
\eea
which vanish at two adjacent roots $y=y_1$ and $y=y_2$ respectively,
each of whose surface gravities is of unit magnitude.  The constants
are therefore given by
\bea
c(i)^{-1} &=& y_i\, (1-\lambda y_i^2) \, 
           \sum_{j=1}^n \fft{1}{y_i^2-a_j^2} - \fft1{y_i}\,,\qquad
\qquad i=1,2\,,\nn\\
b_j(i) &=& -\fft{a_j\, (1-\lambda y_i^2)\, c(i)}{y_i^2-a_j^2}\,,
\qquad\qquad i=1,2\,.
\eea
We shall assume, without loss of generality, that $y_1<y_2$, and we
require that $V(y)-2m>0$ for $y_1 < y < y_2$, to ensure that the
metric has positive definite signature.  With these assumptions, we
shall have
\be
c(1) > 0\,,\qquad c(2) <0\,.
\ee

    The Killing vectors $\ell_1$, $\ell_2$ and $\del/\del\varphi_j$
have zero norm at the degeneration surfaces $r=y_1$ and $r=y_2$
respectively.  The Killing vector $\del/\del\varphi_j$ has zero norm
at the degeneration surface where $\mu_j=0$.  Since the $(n+2)$
Killing vectors $\ell_1$, $\ell_2$ and $\del/\del\varphi_j$ span a
vector space of dimension $(n+1)$, it follows that they must be
linearly dependent.  Since each of the Killing vectors generates a
translation that closes with period $2\pi$ at its own degeneration
surface, it follows that the coefficients in the linear relation must
be rationally related in order not to have arbitrarily nearby points
being identified, and so we may write the linear relation as
\be
p\, \ell_1 + q\, \ell_2 + \sum_{j=1}^n r_j \fft{\del}{\del\varphi_j}=0
\ee
for coprime integers $(p,q,r_j)$.  For the same reason we discussed
for the Einstein-Sasaki cases, here too no subset of $(n+1)$ of these
integers must have any common factor either.  Thus we have the
equations
\be
p\, c(1) + q\, c(2)=0\,,\qquad p \, b_j(1) + q\, b_j(2) + r_j=0\,.
\label{pqsjeqns}
\ee

    Unlike the Einstein-Sasaki limits, however, in this general case
we do not have any relation analogous to (\ref{pqrj}) that imposes a
linear relation on the $(n+2)$ integers.  This is because in the
general case the local metrics (\ref{blodd}) have $(n+1)$ non-trivial
continuous parameters, namely $m$ and the rotations $a_j$, which can
then be viewed as being determined, via the $(n+1)$ equations
(\ref{pqsjeqns}), in terms of the $(n+1)$ rational ratios $p/q$ and
$r_j/q$.  By contrast, in the Einstein-Sasaki limits the local metrics
in $D=2n+1$ dimensions have only $n$ non-trivial parameters, and so
there must exist an equation (namely (\ref{pqrj})) that relates the
$(n+1)$ ratios $p/q$ and $r_j/q$.

    Since, without loss of generality, we are taking $r_1<r_2$, the
integers $p$ and $q$ must be such that $p<q$.\footnote{The case $p=q$,
corresponding to a limit with $r_1=r_2$, was discussed extensively in
\cite{gilupapo1,gilupapo2}, and so for convenience we shall exclude
this from the analysis here.}  From (\ref{pqsjeqns}) we have
\be
r_j = \fft{p\, a_j\, c(1)\, (y_2^2-y_1^2)(1-\lambda a_j^2)}{
           (y_1^2-a_j^2)(y_2^2-a_j^2)}\,.\label{rjlim}
\ee
We must have $y>a_j$ in the entire interval $y_1\le y\le y_2$, in order to
ensure that the metric remains positive definite, and
hence from (\ref{rjlim}) it follows that we must have $r_j>0$.  We shall
denote the associated $D=2n+1$ dimensional 
Einstein spaces by $\Lambda^{p,q,r_1,\ldots, r_n}$.

   From an expression for the determinant of the Kerr-de Sitter metrics
given in \cite{gibperpop}, it is easily seen that the volume of the 
Einstein space $\Lambda^{p,q,r_1,\ldots,r_n}$ is given by
\be
V = \fft{ {\cal A}_{2n+1}}{(\prod_j \Xi_j)}\, 
  \fft{\Delta\tau}{2\pi}\, \Big[ \prod_{j=1}^n 
(y_2^2- a_j^2) - \prod_{j=1}^n(y_1^2-a_j^2)\Big]\, 
\Big( \prod_{k=1}^n \int \fft{d\varphi_k}{2\pi}\Big)\,,
\ee
where ${\cal A}_{2n+1}$ is the volume of the unit $(2n+1)$-sphere.  If
$p$ and $q$ have a greatest common divisor $k=\hbox{gcd}\, (p,q)$ then
the $n$-torus of the $\varphi_j$ coordinates will be factored by a
freely-acting $\Z_k$, and hence it will have volume $(2\pi)^n/k$.  The
period of $\tau$ will be $\Delta\tau=2\pi\, k\, c(1)/q$, and hence the
volume of $\Lambda^{p,q,r_1,\ldots,r_n}$ is
\be
V = \fft{ {\cal A}_{2n+1}}{(\prod_j \Xi_j)}\, 
  \fft{c(1)}{q}\, \Big[ \prod_{j=1}^n 
(y_2^2- a_j^2) - \prod_{j=1}^n(y_1^2-a_j^2)\Big]\,.
\ee

\subsection{Even dimensions}

    A similar discussion applies to the Euclideanised Kerr-de Sitter
metrics in even dimensions $D=2n$, which are also given in
\cite{gilupapo1,gilupapo2}.  There is, however, a crucial difference,
stemming from the fact that while there are $n$ latitude coordinates
$\mu_i$ with $1\le i\le n$, there are only $n-1$ azimuthal coordinates
$\varphi_j$, with $1\le j\le n-1$.  Since the $\mu_i$ coordinates are
subject to the condition $\sum_{i=1}^n \mu_i^2=1$, this means that
now, unlike the odd-dimensional case, there exist surfaces where {\it
all} the azimuthal Killing vectors $\del/\del\varphi_j$ simultaneously
have vanishing norm.  This is achieved by taking
\be
\mu_j=0\,,\qquad 1\le j\le n-1\,;\qquad \mu_n=\pm 1\,.
\ee

    Thus if we consider the Killing vectors $\del/\del\varphi_j$
together with $\ell_1$ whose norm vanishes at $y=y_1$ and $\ell_2$
whose norm vanishes at $y=y_2$, then from the relation
\be
p\, \ell_1 + q\, \ell_2 + \sum_j r_j\, \fft{\del}{\del\varphi_j}=0\,,
\ee
which can be written as
\be
\ell_2 = -\fft{p}{q}\, \ell_1 - 
       \sum_j\fft{r_j}{q}\, \fft{\del}{\del\varphi_j}\,,\label{van2}
\ee
we see that at $(y=y_1,\mu_n=\pm1)$ it will also be the case that
$\ell_2$ has vanishing norm.\footnote{Note that in a positive-definite
metric signature, if two two vectors $A$ and $B$ have vanishing norm
at any point, then so does $A+ \lambda\, B$ for any $\lambda$.  This
can be seen from $(A\pm B)^2\ge0$, which shows that if $A^2$ and $B^2$
vanish at a point, then so does $A\cdot B$.}  Since $\ell_1$, $\ell_2$
and $\del/\del\varphi_j$ all, by construction generate translations at
their respective degeneration surfaces that close with period $2\pi$,
it follows from (\ref{van2}) that there will in general be a conical
singularity at $y=y_2$ associated with a factoring by $\Z_q$. A
similar argument shows there will in general be a conical singularity
at $y=y_1$ associated with a factoring by $\Z_p$.

   The upshot from the above discussion is that one can only get
non-singular Einstein spaces in the $D=2n$ dimensional case if
$p=q=1$.  Since $p=q$, this implies that the two roots $y_1$ and $y_2$
coincide, and hence the analysis reduces to that which was given in
\cite{gilupapo1,gilupapo2}.

   Since the calculations in four dimensions are very simple, it is
instructive to examine this example in greater detail.  The
Euclideanised four-dimensional Kerr-de Sitter metric is given by
\be
ds^2 = \rho^2 \Big(\fft{dy^2}{\Delta_y} +
\fft{d\theta^2}{\Delta_\theta}\Big) +
\fft{\Delta_\theta\sin^2\theta}{\rho^2}\Big(a  d\tau + (y^2-a^2)
\fft{d\phi}{\Xi}\Big)^2 +\fft{\Delta_y}{\rho^2}
\Big(d\tau - a \sin^2\theta \fft{d\phi}{\Xi}\Big)^2\,,
\ee
where
\bea
&&\rho^2=y^2 - a^2 \cos^2\theta\,,\qquad
\Delta_y = (y^2 - a ^2)(1-\lambda\, y^2) - 2m y\,,\nn\\
&&\Delta_\theta =1 - \lambda\, a^2\cos^2\theta\,,\qquad
\Xi = 1 - \lambda a^2\,.
\eea

The function $\Delta_y$ is a quartic polynomial in $y$, which goes to
$-\infty$ for $y\rightarrow \pm \infty$.  Thus one necessary condition
for obtaining a non-singular space is that there exist at least two
real roots.  If there are exactly two real roots, $y_1$ and $y_2$,
with $y_1\le y_2$, then radial variable $y$ must lie in the interval
$y_1\le y\le y_2$.  If there four real roots, then we should choose
$y_1$ and $y_2$ to be adjacent roots, which are either the smallest
pair or the largest pair.

           The Killing vectors that vanish at $y=y_1$ and $y=y_2$
are given by
\be
\ell_i= c_i \fft{\del}{\del\tau} + b_i \fft{\del}{\del\phi}\,,
\ee
where
\be
b_i=\fft{a  (1-a ^2) c_i}{a^2-y_i^2}\,,\qquad
c_i=\fft{2(a^2-y_i^2) y_i}{a ^2 + y_i^2 +
a^2 y_i^2 -3 y_i^4}
\ee
The Killing vector that vanishes at $\sin\theta=0$ is given by
\be
\ell_3=\fft{\del}{\del\phi}\,.
\ee
Thus we have the conditions
\be
p\,\ell_1 + q \ell_2 + r\,\ell_3 = 0
\ee
for $(p,q,r)$ which are pairwise coprime integers.

    For a four-dimensional
compact Einstein space, the Euler number is given by
\be
\chi = \fft1{32\pi^2}\int |\hbox{Riem}|^2\, \sqrt{g}\, d^4 x\,.
\ee
This is easily evaluated for the four-dimensional metrics given above.
With the angular coordinates having periods
\be
\Delta\phi =2\pi\,,\qquad \Delta\tau = \fft{2\pi c_1}{q}\,,\label{periods}
\ee
we find that
\be
\chi = \fft2{p} + \fft2{q}\,.\label{chires}
\ee
If $p=q=1$ we have $\chi=4$. This is indeed the correct Euler number
for the $S^2$ bundle over $S^2$, which, as shown in \cite{pagemet} and
\cite{gilupapo1,gilupapo2}, is the only non-singular case that arises
when the roots $y_1$ and $y_2$ coincide.  If one were to consider
cases where $p\ne 1$ or $q\ne 1$, then in general $\chi$ would not be
an integer, in accordance with our observations above that in such
cases there are $\Z_p$ and $\Z_q$ orbifold type singularities in the
space.  It is possible that one might be able to blow up these
singularities, and thereby obtain a non-singular space with a more
complicated topology.  The fact that the ``Euler number'' $\chi$ given
in (\ref{chires}) has a simple rational form can perhaps be taken as
supporting evidence.

\section{Conclusions}

   In this paper, we have elaborated on the results which we obtained
in \cite{cvlupapo}, constructing new Einstein-Sasaki spaces
$L^{p,q,r_1, \cdots,r_{n-1}}$ and non-Sasakian Einstein spaces
$\Lambda^{p,q,r_1,\cdots, r_n}$, in all odd dimensions $D=2n+1\ge 5$.
These spaces are all complete and non-singular compact manifolds.  The
metrics have cohomogeneity $n$, with isometry group $U(1)^{n+1}$,
which acts transitively on the $(n+1)$-dimensional principal orbits.

    The Einstein-Sasaki metrics arise after Euclideanisation, by
taking certain BPS limits of the Kerr-de Sitter spacetimes constructed
in $D=5$ in \cite{hawhuntay}, and in all higher dimensions in
\cite{gilupapo1,gilupapo2}.  The BPS limit effectively implies that
there is a relation between the mass and the $n$ rotation parameters
of the $(2n+1)$-dimensional Kerr-de Sitter metric, and thus the local
Einstein-Sasaki metrics have $n$ non-trivial free parameters.  These
metrics are in general singular, but by imposing appropriate
restrictions on the parameters, we find that the metrics extend
smoothly onto complete and non-singular compact manifolds, which we
denote by $L^{p,q,r_1,\cdots,r_{n-1}}$, where the integers
$(p,q,r_1,\ldots, r_{n-1})$ are coprime.

    In the case of the five-dimensional Einstein-Sasaki spaces
$L^{p,q,r}$, we have been able to obtain an explicit formula
expressing the volume in terms of the coprime integers $(p,q,r)$, via
a quartic polynomial.  In the AdS/CFT correspondence, it is expected
that the boundary field theory dual to the type IIB string on
AdS$_5\times L^{p,q,r}$ will be a quiver gauge theory. In particular,
the central charge of the quiver theory should be related to the
inverse of the volume of $L^{p,q,r}$. The central charges have
recently been calculated using the technique of $a$-maximisation, and
it has been shown that they are indeed precisely in correspondence
with the volumes given by our polynomial (\ref{quartic})
\cite{benkru}.  (See \cite{martspar2,berbigcot,befrhamasp,marspayau}
for the analysis of the dual quiver theories, and $a$-maximisation,
for the previous $Y^{p,q}$ examples.)

    We have also shown, for the five-dimensional $L^{p,q,r}$ spaces,
that a convenient characterisation can be given in terms of the two
parameters $u$ and $v$, introduced in equation (\ref{uv1}), where
$0<u<v<1$.  If $u$ and $v$ are taken to be arbitrary rational numbers
in this range, then we obtain a corresponding Einstein-Sasaki space
that is ``quasi-regular,'' meaning that the orbits of $\del/\del\tau$
are closed.  In general, when $u$ and $v$ are irrational numbers,
again related to the coprime integers $(p,q,r)$ by (\ref{uv1}), the
orbits of $\del/\del\tau$ will never close, and the corresponding
Einstein-Sasaki space is called ``irregular.''\footnote{One should not
be misled by the terminology ``quasi-regular'' and ``irregular'' that
is applied to Einstein-Sasaki spaces; all the spaces $L^{p,q,r}$ are
complete and non-singular.}

   Several other papers have appeared making use of our results in
\cite{cvlupapo}. These include \cite{geppal}, where branes in
backgrounds involving the $L^{p,q,r}$ spaces are constructed, and
\cite{ahnpor}, where marginal deformations of eleven-dimensional
backgrounds involving the $L^{p,q,r_1,r_2}$ spaces are constructed.

   In addition to discussing the Einstein-Sasaki spaces
$L^{p,q,r_1,\cdots, r_{n-1}}$, we have also elaborated in this paper
on the non-Sasakian Einstein spaces $\Lambda^{p,q,r_1,\cdots,r_n}$
that were constructed in $D=2n+1$ dimensions in \cite{cvlupapo}.
These arise by Euclideanising the Kerr-de Sitter metrics without
taking any BPS limit.  As local metrics they are characterised by
$(n+1)$ parameters, corresponding to the mass and the $n$ independent
rotations of the $(2n+1)$-dimensional rotating black holes. Again, by
choosing these parameters appropriately, so that they are
characterised by the $(n+2)$ coprime integers $(p,q,r_1,\ldots, r_n)$,
we find that the local metrics extend smoothly onto complete and
non-singular compact manifolds.

\bigskip
  
\noindent{\Large{\bf Acknowledgements}}

   M.C. and D.N. Page are grateful to the George P. \& Cynthia
W. Mitchell Institute for Fundamental Physics for hospitality.
Research supported in part by DOE grants DE-FG02-95ER40893 and
DE-FG03-95ER40917, NSF grant INTO3-24081, the NESRC of Canada, the
University of Pennsylvania Research Foundation Award, and the Fay
R. and Eugene L.Langberg Chair.

\bigskip
\bigskip\bigskip

\noindent{\Large \bf Appendix}

\appendix

\section{Singular Limits}

    In this paper we have focussed on limits of the Euclideanised Kerr
de Sitter metrics in odd dimensions, in which all the rotation
parameters $a_i$ are subjected to the limiting procedure given in
equation (\ref{dbpslim}).  As we have seen, this yields a rich source
of non-singular Einstein-Sasaki spaces, which we denote by
$L^{p,q,r_1,\cdots,r_{n-1}}$ in $D=2n+1$ dimensions.

    In fact, a BPS limit also arises if only a subset of the rotation
parameters are subjected to the limiting procedure given in
(\ref{dbpslim}).  Here, we study the resulting metrics, and we show
that they are singular in all such cases.

   After performing the Euclideanisation as in section \ref{esnsec},
we now take limits as follows:
\bea &&a_i=\lambda^{-\ft12} (1 - \ft12\alpha_i\,\epsilon)\,,
    \ \ i=1,\ldots, p\le n
\nn\\
&&r^2=\lambda^{-1} (1 - x\epsilon)\,,\quad m=\ft12\lambda^{-1} \mu
\epsilon^{p+1} 
\eea
and then send $\epsilon\rightarrow 0$.  Note that for $p<n$, the
angular momentum parameters $a_{j}<\lambda^{-\ft12}$ with
$j=p+1,\ldots,n$ are not scaled.  This limit corresponds to Lorentzian
BPS solutions with
\be 
E= g\sum_{i=1}^p J_i\,, 
\ee
and has more supersymmetry than in the $p=n$ case that we discussed in 
section \ref{esnsec}.  The metrics (\ref{blodd}) become
\be 
\lambda\,ds_{2n+1}^2 = (d\tau + \sigma)^2 + ds_{2n}^2\,,\label{Dmet} 
\ee
where
\bea 
d\bar s_{2n}^2 &=& \fft{Y dx^2}{4x F} - \fft{x(1-F)}{Y}\Big(\sum_{i=1}^p
\alpha_i^{-1}\,
 \mu_i^2 d\varphi_i\Big)^2 + 
\sum_{i=1}^p (1-\alpha_i^{-1}\, x)(d\mu_i^2 + \mu_i^2 d\varphi_i^2)
\nn\\
&&+ \fft{x}{\sum_{i=1}^p \alpha_i^{-1} \mu_i^2}\,
        \Big( \sum_{j=1}^p \alpha_j^{-1}\, \mu_j d\mu_j\Big)^2\,
- \sigma^2\
        + \sum_{k=p+1}^n (d\mu_k^2 + \mu_k^2 d\varphi_k^2) \nn\\
&&+ \sum_{i=1}^p (1-\alpha_i^{-1}\, x)(d\mu_i^2 + \mu_i^2 d\varphi_i^2)
\nn\\
\sigma &=& \sum_{i=1}^p (1-\alpha_i^{-1}x)\mu_i^2\, d\varphi_i\,,
\label{genmet}
\eea
where $\sum_{i=1}^n \mu_i^2=1$, and
\be 
Y=\sum_{i=1}^p\fft{\mu_i^2}{\alpha_i-x}\,,\qquad
  F= 1- \fft{\mu}{x}\, \prod_{i=1}^p(\alpha_i-x)^{-1}\,.
\ee
Note that in this scaling limit all the dependence on $a_j$ for
$j=p+1,\ldots, n$ has been absorbed into a redefinition of the mass
parameter $\mu$.  For $p<n$, one can introduce a new set of
coordinates:
 \bea   \mu_i &=&
\nu_i \, \sin\gamma\,, \ \ i=1, \ldots, p\, , \nn\\
\mu_{j+p}& =& \td\mu_j\, \cos\gamma \, , \ \ j=1,\ldots, (n-p)\,, 
\eea
where $\sum_{i=1}^p \nu_i^2=1$ and $\sum_{j=1}^{n-p}{\tilde
\mu}_j^2=1$. The metric can now be cast in the form:
\be 
ds_{2n+1}^2= d\gamma^2 +\cos^2{\gamma}\, d
\Omega_{2n-2p-1}^2+\sin^2\gamma \, dS_{2p+1}^2\, . \label{conemetric}
\ee
Here $d\Omega_{2n-2p-1}^2$ is the metric on the unit
$(2n-2p-1)$-sphere, and $dS_{2p+1}^2$ is the previously-obtained
Einstein-Sasaki metric (\ref{esgenmet}) in $2p+1$ dimensions with all
$\alpha_i\ne 0$ ($i=1\cdots p$).  Note that since $dS_{2p+1}^2$ is an
Einstein-Sasaki space (and not a round sphere), the metric
(\ref{conemetric}) is singular at $\gamma=0$. A cone over
(\ref{conemetric}) produces a direct product of spaces in $D=2n+2$,
namely ${\bf R}^{2(p-n)}\times C_{2p+2}$, where $C_{2p+2}$ is a $2p+2$
dimensional Calabi-Yau cone.

   The above discussion was for the case of BPS limits for the Kerr-de
Sitter metric in odd dimension $D=2n+1$, in which only $p$ of the $n$
rotation parameters were subjected to the limiting procedure in
(\ref{dbpslim}).  A very similar discussion can be given for BPS
limits of the Kerr-de Sitter metrics in even dimension $D=2n$, whose
metrics are also given in \cite{gilupapo1,gilupapo2}.  In this case
there are again $n$ coordinates $\mu_i$, but only $(n-1)$ azimuthal
angles $\varphi_i$, and so the situation is similar to the
odd-dimensional case when one of the rotation parameters is set to
zero.  For this reason, we find that all BPS limits give rise to
singular metrics, similar in form to (\ref{conemetric}).  In fact,
applying the limiting procedure in (\ref{dbpslim}) to all $(n-1)$
rotations, we obtain the metric
\be 
ds_{2n}^2= d\gamma^2 +\sin^2\gamma \, dS_{2n-1}^2\,,
\label{conemetriceven}
\ee
where $dS_{2n-1}^2$ is an Einstein-Sasaki metric of the kind obtained
in section \ref{esnsec}.  Clearly(\ref{conemetriceven}) is singular at
$\gamma=0$, and the metric does not extend onto a non-singular
manifold.

\end{document}